# EVALUACIÓN EXPERIMENTAL DEL FACTOR DE ESTABILIDAD EN NANOEMULSIONES DODECANO/AGUA

Yorlis Mendoza, Kareem Rahn-Chique, German Urbina-Villalba*

Instituto Venezolano de Investigaciones Científicas (IVIC), Centro de Estudios Interdisciplinarios de la Física (CEIF), Laboratorio de Fisicoquímica de Coloides, Carretera Panamericana Km. 11, Altos de Pipe, Edo. Miranda, Apdo. 20632, Caracas, Venezuela. e-mail: guv@ivic.gob.ve

**Resumen** Se estudia la variación temporal de la turbidez de cuatro nano-emulsiones iónicas de aceite en agua (o/w) estabilizadas con dodecil-sulfato de sodio (DSS), compuestas por dodecano puro, y mezclas de dodecano con escualeno y tetracloroetileno. Para cada sistema se reporta la curva del factor de estabilidad en función de la fuerza iónica del medio, y se discute la importancia relativa de la flotabilidad y la maduración de Ostwald sobre la evaluación de las tasas de agregación.

**Palabras clave** Nanoemulsión, Agregación, Ostwald, W, Flotabilidad.

## 1. INTRODUCCIÓN

Hace casi 100 años Smoluchowski (1917) desarrolló una teoría para la agregación irreversible de partículas sólidas de igual tamaño suspendidas en un líquido, cuyo movimiento es producto –exclusivamente– de su interacción térmica con el solvente. De acuerdo con Smoluchowski, el movimiento Browniano de las partículas promueve colisiones que generan su unión irreversible. La concentración numérica de flóculos de k partículas existentes en la dispersión luego de transcurrido un tiempo t, $n_k(t)$, resulta de la competencia entre el número de flóculos de tamaño k producido por el choque de agregados más pequeños de tamaño i y j tal que i+j = k, y el número de flóculos de tamaño k que se pierde por colisiones con agregados de cualquier otro tamaño:

$$\frac{dn_k(t)}{dt} = \frac{1}{2}\sum_{i=1\ j=k-i}^{i=k-1} k_{ij} n_i(t) n_j(t) - n_k(t)\sum_{i=1}^{\infty} k_{ik} n_i(t)$$
$$k = 1, 2, 3... \quad (1)$$

La ecuación (1) se conoce con el nombre de balance de poblaciones de la dispersión. El kernel de la ecuación es el conjunto de constantes de floculación entre agregados de tamaño i y j: $\{k_{ij}\}$. La evaluación de éstas constantes es difícil, por lo que generalmente la velocidad de agregación se estima a través del valor de la tasa de formación de dobletes, $k_{11}$, la cual puede evaluarse mediante medidas de turbidez. Por definición la turbidez de una dispersión ($\tau_{exp}$) es igual a:

$$\tau_{exp} = (1/L_c)\ln(I_0/I) \quad (2)$$

Donde $I_0$ es la intensidad de luz incidente, e I es la intensidad de luz emergente de una celda de ancho $L_c$ (generalmente del orden de $10^{-2}$ m). Teóricamente la turbidez puede expresarse como el producto de la densidad de partículas de cada tamaño existentes en un tiempo t por su sección transversal óptica ($\sigma_k$):

$$\tau_{theo} = \sum_{k=1}^{\infty} n_k \sigma_k \quad (3)$$

En el límite de tiempos muy cortos con respecto al tiempo de vida media de la dispersión, $\tau_{1/2} = (k_{11}n_0/2)^{-1}$, la turbidez es fundamentalmente causada por las partículas presentes inicialmente en la solución (singletes) y los dobletes formados por ellas durante la primera etapa de la agregación. En estas condiciones puede deducirse que [Lips, 1971; Lips, 1973]:

$$\left(\frac{d\tau}{dt}\right)_0 = \left[\frac{\ln 10}{L_c}\right]\left(\frac{dAbs}{dt}\right)_0 = \left(\frac{1}{2}\sigma_2 - \sigma_1\right) k_{11} n_0^2 \quad (4)$$

Donde Abs = $\log(I_0/I)$ es la absorbancia de la dispersión, $n_0 = n_a(t=0)$ es el número total de agregados al comienzo del proceso, y $\sigma_1$ y $\sigma_2$ son las secciones transversales ópticas de singletes y dobletes, respectivamente.

Derjaguin, Landau, Verwey y Overbeek (DLVO) [Verwey, 1946] demostraron que la difusión de las partículas en un solvente está influenciada por las interacciones entre las mismas, las cuales suelen ser de dos tipos: a) Atractivas, debido a la interacción dipolar entre las moléculas que las componen, y b) Repulsivas, debido a la carga superficial de las mismas. En el caso de emulsiones de aceite en agua estabilizadas con dodecil sulfato de sodio, la interacción atractiva depende del tipo de aceite y la carga de las gotas es el producto de la adsorción del surfactante a sus superficies.

Si la carga superficial de las gotas se "apantalla" utilizando una concentración de sal suficientemente alta, la tasa de floculación $k_{11}$ que puede evaluarse empleando la ecuación (4) es una tasa de agregación rápida. Si en cambio, la salinidad del medio es moderada ó inexistente, el sistema es relativamente estable con respecto a su comportamiento a alta salinidad, y su tasa de agregación es lenta. En consecuencia, puede definirse un factor de estabilidad (W) evaluando la variación temporal de la turbidez en condiciones de agregación rápida y lenta.

$$W = \left(\frac{dAbs}{dt}\right)_0^{rápida} \Big/ \left(\frac{dAbs}{dt}\right)_0^{lenta} = k_{11}^{rápida} \Big/ k_{11}^{lenta} \quad (5)$$

La ecuación (5) evita la evaluación de las secciones transversales ópticas de la ecuación (4) pero sólo permite estimar la estabilidad relativa de una dispersión con respecto a su tasa de agregación rápida. En trabajos recientes [Rahn-Chique, 2012a; 2012b], nuestro grupo ha calculado tanto los valores de $k_{11}$ como los de W para emulsiones de dodecano en agua estabilizadas con dodecil sulfato de sodio. Sin embargo sólo recientemente se ha estudiado el efecto que la maduración de Ostwald y la flotabilidad de las gotas tiene sobre las tasas de floculación [Mendoza, 2012].

La maduración de Ostwald consiste en el intercambio difusivo de moléculas de aceite entre gotas de aceite de distinto tamaño a través del agua. Este intercambio se debe a la diferencia de potencial químico entre dichas gotas. A la fecha, la teoría más importante de maduración de Ostwald es la teoría LSW [Lifshitz, 1961; Wagner, 1961], la cual predice una variación inversamente proporcional entre el número total de partículas suspendidas en solución y el tiempo:

$$n_g = \sum_i n_{g,i} = \left[\frac{1}{2\alpha D_m C_\infty}\right]\frac{1}{t} \quad (6)$$

Donde $n_{g,i}$ es el número de *gotas* de tamaño i existente en el tiempo t. Nótese que en la Ec. (6), el subíndice i corre sobre las gotas de cada tamaño y no sobre el número de agregados de gotas.

Adicionalmente, LSW predice un incremento lineal del radio cúbico promedio en función del tiempo:

$$V_{OR} = dR_p^3 \big/ dt = 4\alpha D_m C(\infty)/9 \quad (7)$$

En la Ec. (7) $R_p$, $D_m$, $C(\infty)$ y $\alpha$ representan el radio promedio, la constante de difusión de las moléculas de aceite en agua, la solubilidad del aceite en presencia de una interfase plana, y la longitud capilar:

$$\alpha = 2\gamma V_M \big/ \tilde{R}T \quad (8)$$

Donde $\tilde{R}$ es la constante universal de los gases, $V_m$ el volumen molar del aceite, y T la temperatura absoluta.

La definición de la fracción de volumen de aceite ($\phi$) conjuntamente con las ecuaciones (6) y (7) permiten encontrar la dependencia $n_g(t)$ en función de la condición inicial del sistema. Así [Urbina-Villalba, 2012a]:

$$V_{OR} = \frac{dR_p^3}{dt} = \frac{3\phi}{4\pi}\frac{d}{dt}\left(\frac{1}{n_g}\right) =$$
$$\frac{3\phi}{4\pi}\left(-\frac{1}{n_g^2}\right)\frac{dn_g}{dt} = \frac{4\alpha D_m C_\infty}{9} \quad (9)$$

Por tanto:

$$\frac{dn_g}{dt} = -\left[\frac{4\alpha D_m C_\infty}{9}\right]\left[\frac{4\pi}{3\phi}\right]n_g^2 \quad (10)$$

Esto nos lleva a deducir una ecuación simple para la variación del número total de gotas producto de la maduración de Ostwald [Urbina-Villalba, 2012a]:

$$n_g(t) = \frac{n_0}{1 + k_O n_0 t} \quad (11)$$

La ecuación (11) permite definir una tasa de maduración de Ostwald en función de la variación del número total de gotas que existen en el tiempo t:

$$k_O = \frac{16\pi\alpha D_m C_\infty}{27\phi} \quad (12)$$

Nótese que la ecuación (11) es similar a la variación del número total de agregados en función del tiempo predicha por Smoluchowski para el caso en que las tasas de agregación entre flóculos de tamaño distinto son muy similares ($k_{ij} \approx k_f \approx k_{11}/2$):

$$n_a = \frac{n_0}{1 + k_F n_0 t} \tag{13}$$

Donde $n_a$ es el número total de agregados existentes en la dispersión en el tiempo t:

$$n_a = \sum_k n_k = \sum_k \left[ \frac{n_0 (k_F n_0 t)^{k-1}}{(1 + k_F n_0 t)^{k+1}} \right] \tag{14}$$

La similitud entre las ecuaciones (11) y (13) hacen preguntarse si el fenómeno de maduración de Ostwald influye en la evaluación de $k_{11}$, ya que este fenómeno promueve el cambio de la distribución de tamaño de gotas como función del tiempo. Curiosamente, y al menos dentro de las limitaciones de la teoría LSW, la tasa de maduración no cambia como producto de la floculación. Sin embargo, sí es afectada por la coalescencia que pudiese ocurrir a consecuencia de la agregación, ya que ésta última afecta el tamaño promedio de partículas en suspensión.

Por otra parte es bien sabido que las gotas de una emulsión experimentan una fuerza de flotación constante, producto de la competencia entre el efecto gravitatorio y la ley de Arquímedes:

$$F_b = 4\pi R_i^3 (\rho_o - \rho_w) g / 3 \tag{15}$$

Esta fuerza causa la formación de crema en el tope del recipiente y genera un gradiente de concentración en el número de partículas por unidad de volumen. En ausencia de agregación, la velocidad de formación de crema puede estimarse a partir del término convectivo de la ecuación de movimiento Browniano [Urbina-Villalba, 2009a]:

$$\Delta L = \frac{D_i F}{k_B T} \Delta t \Rightarrow$$
$$V_g = \frac{\Delta L}{\Delta t} \approx \frac{[k_B T / 6\pi \eta R_i] F_b}{k_B T} \tag{16}$$

Donde $\Delta L$ es la altura del recipiente que contiene la emulsión, $\Delta L = r_i(t+\Delta t) - r_i(t)$, $r_i(t)$ es la posición de la gota "i" en el tiempo t, $\eta$ la viscosidad de la fase acuosa, y $k_B$ la constante de Boltzmann. De acuerdo a las Ecs. (15) y (16) la fuerza de flotación es igual a:

$$V_g = \frac{[k_B T / 6\pi \eta R_i][4\pi R_i^3 \Delta \rho g]}{3 k_B T} = \frac{2 R_i^2 \Delta \rho g}{9\eta} \tag{17}$$

La velocidad de formación de crema afecta las medidas de agregación en al menos dos formas. Por una parte, la absorbancia captada por el equipo cambia si las gotas migran por flotación fuera de la zona de muestreo del láser incidente. Por otra parte, las constates de agregación se modifican en presencia de una fuerza convectiva. Esto se conoce como agregación pericinética, la cual es generalmente más rápida que la agregación resultante de un proceso de difusión (agregación ortocinética). A fin de comprender estas diferencias es útil imaginarse una gota grande y una gota pequeña ubicadas en el eje vertical del recipiente, una encima de la otra. Si la grande se sitúa debajo de la pequeña, la flotabilidad inducirá la floculación de ambas. Si la gota grande se sitúa arriba de la pequeña, ambas tenderán a separarse con el transcurrir del tiempo y no flocularán. En promedio se espera que la fuerza de flotación incremente la velocidad de floculación debido al aumento del movimiento relativo entre las gotas.

En ausencia de agregación la tasa de formación de crema de nanoemulsiones es despreciable. Así una gota de dodecano con un radio de 100 nm situada en el fondo de un recipiente de 10 cm de altura, requeriría 205 días en llegar al tope. A pesar de ello, con frecuencia se observa que una nanoemulsión iónica forma crema en 3 días, lo cual evidencia la ocurrencia de agregación y/ó coalescencia y/ó maduración de Ostwald.

Se conoce [Kamogawa, 1999] que la adición de un componente insoluble al aceite disminuye su solubilidad. Por otra parte también es sabido que la mezcla de un aceite menos denso con otro más denso, produce una combinación de densidad intermedia. Esto hace posible la generación de mezclas de flotabilidad nula, adicionando tetracloroetileno (TCE) al dodecano.

En este trabajo se estudia la velocidad de agregación relativa de cuatro emulsiones a través de la evaluación de su cociente de estabilidad. Estas emulsiones están formadas por: a) dodecano (Nano A), b) dodecano ($C_{12}$) + escualeno (SQ) (Nano B), c) $C_{12}$ + TCE, (Nano C), y finalmente, d) $C_{12}$ + SQ + TCE (Nano D). El primer sistema (A) es sus-

Yorlis Mendoza, Kareem Rahn-Chique, German Urbina-Villalba

ceptible a todos los mecanismos de desestabilización. En los sistemas B y C se ralentiza la velocidad de maduración de Ostwald y la flotabilidad, respectivamente. Con el sistema D se intenta corregir simultáneamente la maduración de Oswald y la flotabilidad de la emulsión original compuesta por dodecano únicamente. De esta manera se puede estimar el efecto del intercambio molecular y la gravedad sobre la tasa de agregación de una nanoemulsión de dodecano.

## 2. PROCEDIMIENTO EXPERIMENTAL

### 2.1. Materiales

El dodecano (Aldrich, 99%, 0,75 g/cc) se eluyó dos veces a través de una columna de alumina a fin de incrementar su pureza. El cloruro de sodio (Sigma, 99,5%), el iso-pentanol (Scharlau Chemie, 99%, 0,81 g/cc), el tetracloroetileno (J.T. Baker, 100%, 1,614 g/cc), el dodecil sulfato de sodio (Sigma, 99%) y el escualeno (Aldrich, 99%, 0,809 g/cc) se usaron tal y como fueron recibidos. El agua fue destilada y deionizada (1,1$\mu$S/cm$^{-1}$ at 25ºC) usando un purificador Simplicity de Millipore (USA).

### 2.2. Síntesis y Caracterización de las Nanoemulsiones

Las nanoemulsiones se prepararon por el método de inversión de fases por composición (Solè *et al.*, 2006; Wang *et al.*, 2008). A objeto de garantizar la ocurrencia de mínima tensión durante el mezclado de los componentes, se construyó previamente un diagrama de fases del sistema compuesto por dodecano, NaCl, iso-pentanol, dodecil sulfato de sodio (DSS) y agua (Rahn-Chique, 2012a; Rahn-Chique, 2012b). Para las emulsiones de aceites mezclados se empleó un protocolo similar cambiando levemente el modo de adición de agua y la forma de agitación.

La cantidad de TCE necesaria para producir una mezcla de flotación nula (Nano C) se estimó inicialmente aproximando la densidad de la mezcla por la suma de las densidades de los componentes puros multiplicada por su fracción másica en la mezcla. Las pruebas de flotación se realizaron primeramente en un beaker mediante agitación manual, y luego en un tubímetro de altura variable (Quickscan, Beckman-Coulter). La cantidad de escualeno necesaria para producir una emulsión estable respecto a maduración de Ostwald (Nano B) se estimó incrementando la proporción de SQ en C$_{12}$ sistemáticamente, y monitoreando la variación del radio cúbico promedio con respecto al tiempo. Para la Nano D fue necesario establecer una composición de compromiso entre la flotación y la maduración de Ostwald, la cual se logró con la siguiente composición: C$_{12}$ 56 % wt/wt, TCE 23% wt/wt, SQ 21% wt/wt (Nano D).

Para estudiar la velocidad de agregación de los sistemas antes mencionados, se empleó una concentración numérica de gotas por unidad de volumen igual a $n_0 = 4 \times 10^{10}$ part/cm$^3$. La concentración de DSS se ajustó durante la dilución a 8 mM. Para la selección de la longitud de onda apropiada, el espectro de absorción de cada componente y de las nanoemulsiones preparadas fue estudiado entre 200 y 1100 nm. Se escogió aquella longitud de onda a la cual los componentes de las mezclas y las nanoemulsiones preparadas no absorbían la luz de manera sustancial ($\lambda$= 800 nm). La concentración inicial de partículas se halló realizando un barrido de $(dAbs/dt)_0$ vs. $n_0$ en condiciones de agregación rápida ([NaCl] = 600 mM). La pendiente $(dAbs/dt)_0$ se refiere a la variación inicial de la curva de Absorbancia vs. tiempo de una emulsión dada.

Para las medidas de absorbancia en función de [NaCl] para cada emulsión se empleó una concentración de sal entre 350 y 600 mM. En cada caso una alícuota de emulsión madre fue diluida a la concentración de trabajo. Luego, 0,6 ml de solución salina concentrada fueron añadidos a 2,4 ml de emulsión diluida. La turbidez fue detectada durante 60 segundos con un espectrofotómetro Turner. Las medidas se repitieron un mínimo de tres veces a cada concentración. La Figura 1 ilustra las curvas típicas de Absorbancia vs. tiempo obtenidas en este tipo de experimento. Como se observa en el recuadro inferior derecho, al principio la absorbancia es constante hasta que se abre la compuerta de la celda de medida y se introduce la solución salina. Debido a la dilución la absorbancia inicial baja. Durante un tiempo finito su variación es errática producto de la falta de homogeneidad del medio y la perturbación producida por la luz externa. Seguidamente la absorbancia se incrementa de forma monótona como producto de la formación de agregados. El comienzo de esta curva corresponde a la formación de dobletes. De allí que su pendiente inicial sea proporcional a $k_{11}$ (Ec. (4)). El cociente de estabilidad fue evaluado a partir de la Ec. (5). Para ello se estimó que la agregación rápida podía alcanzarse a 600 mM NaCl.

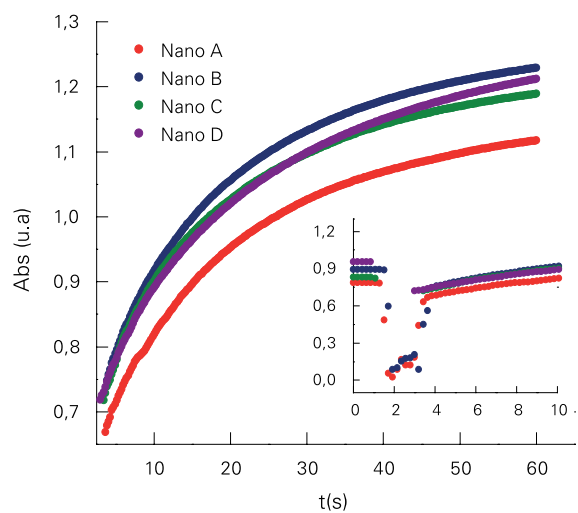

Figura 1: Curvas típicas de Absorbancia vs. tiempo para cada nanoemulsión en el régimen DLCA ([NaCl] = 600 mM).

De acuerdo con Reerink y Overbeek [1954] la pendiente de las gráficas de Log W vs. Log [NaCl] permite la evaluación de la constante de Hamaker de las partículas interactuantes, y de su potencial eléctrico. Dichas ecuaciones son:

$$A_H = \sqrt{\frac{1{,}73 \times 10^{-57} \, (d\log W / d\log[NaCl])^2}{R^2 \, (CCC)}} \quad (18)$$

$$\Psi_0 = 0{,}1028 \; \text{Arc} \; \tanh \sqrt{\frac{(d\log W / d\log[NaCl])^2}{R(2.15 \times 10^9)}} \quad (19)$$

En estas ecuaciones las constantes son tales que el radio de las partículas (R) debe expresarse en metros, y la concentración de coagulación crítica (CCC) en moles por litro. Así la constante de Hamaker se obtiene en Joules, y el potencial superficial en Voltios.

Las Ecs. (18) y (19) son ecuaciones aproximadas deducidas empleando una fórmula para el potencial repulsivo que sólo es válida para potenciales eléctricos superficiales inferiores a 25,6 mV ($\approx k_B T$, para T = 298 K). Tal y como se muestra en la Tabla 1, el potencial eléctrico de las gotas de las emulsiones A, B, C, D es muy superior a este valor.

### 3. RESULTADOS Y DISCUSIÓN

Algunas propiedades de la nanoemulsiones preparadas pueden verse en la Tabla 1. Dado que el potencial eléctrico repulsivo entre las partículas disminuye a medida que aumenta la fuerza iónica del medio, cabe esperar que W disminuya hasta alcanzar el valor de 1,0 alrededor de la concentración de coagulación crítica (CCC). Este es el comportamiento se ilustra en las Figuras (2) – (5), siendo el valor

Tabla 1: Propiedades de las Nanoemulsiones

| Sistema | Potencial Zeta (mV) | Diámetro Promedio (nm) | Tiempo en días necesario para ascender 10 cm (Exp.) | Tiempo estimado para ascender 10 cm (Teórico) |
|---|---|---|---|---|
| A | -83,8 | 435 | 32 | 43 |
| B | -88,6 | 416 | 55 | 49 |
| C | -72,6 | 413 | ∞ | 19817 |
| D | -93,8 | 402 | 340 | 342 |

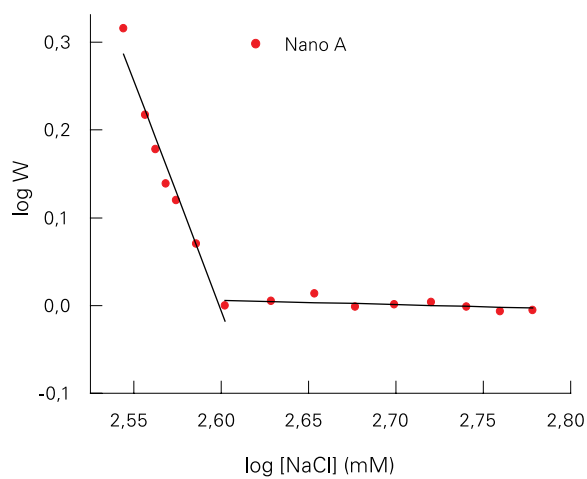

Figura 2: Log W vs. Log [NaCl] para Nanoemulsiones Tipo A.

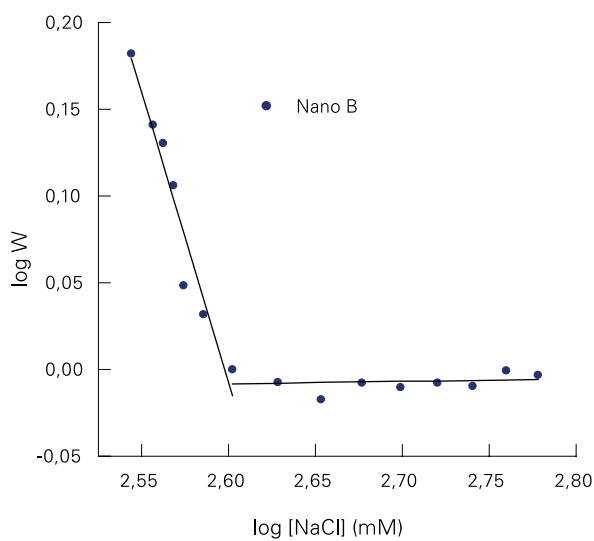

Figura 3: Log W vs. Log [NaCl] para Nanoemulsiones Tipo B.

Yorlis Mendoza, Kareem Rahn-Chique, German Urbina-Villalba

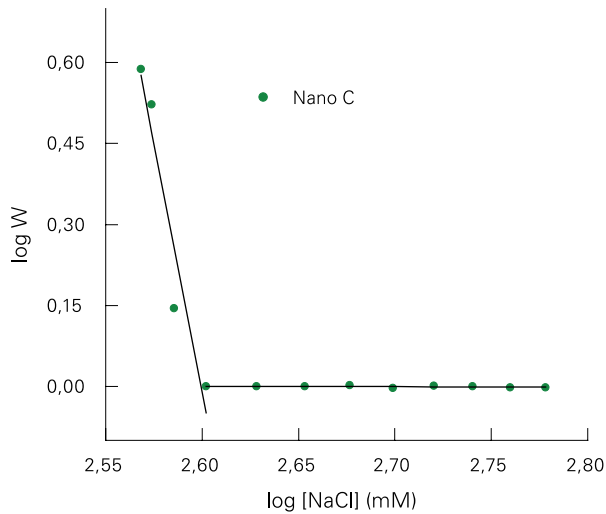

Figura 4: Log W vs. Log [NaCl] para Nanoemulsiones Tipo C.

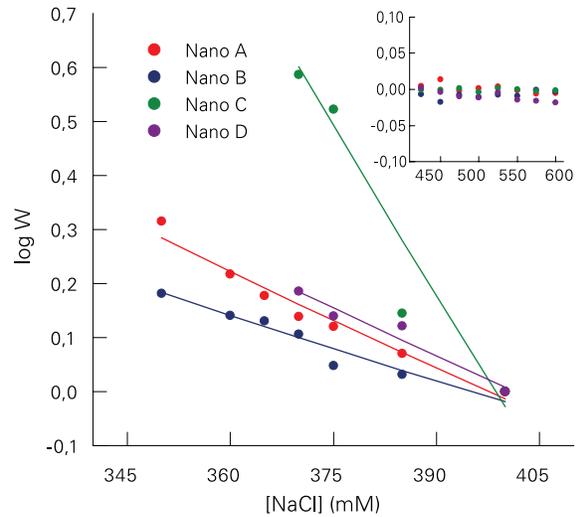

Figura 6: Log W vs. [NaCl] para el régimen RLCA.

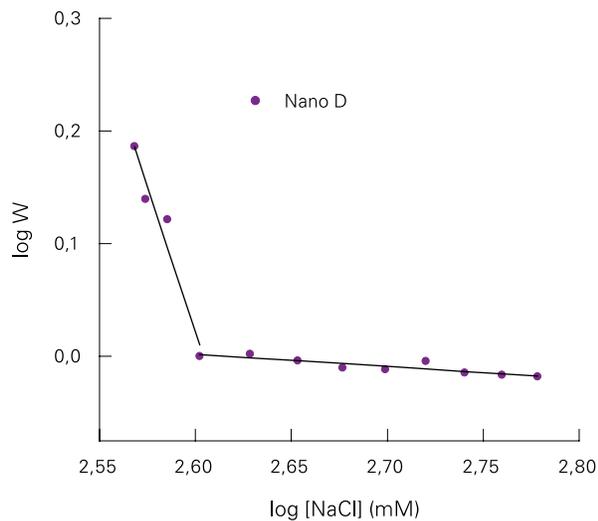

Figura 5: Log W vs. Log [NaCl] para Nanoemulsiones Tipo D.

de la CCC en todos los casos 400 mM. El potencial superficial de éstos sistemas medido por electroforesis aumenta en el orden: C < A < B < D. En consecuencia cabría esperar que para las mismas salinidades los valores del cociente de estabilidad W varíen en el orden $W_D > W_B > W_A > W_C$. Curiosamente este no es el caso (ver Figura 6). Se encuentra que $W_C > W_D > W_A > W_B$. Esta variación si está de acuerdo con los valores del potencial eléctrico superficial que se deducen a partir de la ecuación (19), a pesar de que la magnitud de dichos valores es extremadamente baja en comparación con los potenciales evaluados experimentalmente (Tabla 1). Tal y como se mencionó en la sección anterior, la fórmula del potencial repulsivo empleado por Reerink y Overbeek es inadecuada para potenciales eléctricos grandes. En consecuencia no es sorprendente que los valores generados por la Ec. (19) discrepen sustancialmente de los valores medidos. En este sentido debe observarse que las curvas de Log W vs. Log [NaCl], y las de Log W vs. [NaCl], muestran coeficientes de regresión similares, lo cual sugiere que otra forma analítica del potencial eléctrico que conduzca a una relación lineal de Log W vs. [NaCl] es posible (Tabla 2).

Cabe mencionar que la estabilidad de los sistemas C y D no pudo ser medida por debajo de 370 mM, lo cual crea una incertidumbre grande en el valor de la pendiente de Log W vs. Log [NaCl]. El sistema D mostró emulsiones completamente estables por debajo de 370 mM: las dispersiones lucían homogéneas luego de la medida y se observó que su absorbancia no cambiaba como función del tiempo. En la misma región de salinidades ([NaCl] < 370) el sistema C mostró curvas de absorbancia que inicialmente subían pero luego bajaban originando una especie de máximo ancho. Dado que el origen de tal variación es incierto especialmente en un sistema que tiene la densidad de las fases igualada, se prefirió dejar los datos de las pendientes de tales curvas fuera de este estudio.

La estabilidad de los sistemas con respecto a formación de crema en ausencia de sal fue determinada por medidas de turbidez en función de la altura. De allí se calculó la tasa de formación de crema y se estimó el tiempo necesario para recorrer la longitud total de la celda de medida (10 cm). Estos resultados se muestran en la Tabla 1. Allí también se presentan los estimados obtenidos empleando la fórmula (17). Debe notarse que pese a nuestros esfuerzos no fue posible obtener emulsiones con el mismo radio promedio para los diferentes sistemas. Sin embargo las diferencias se encuentran dentro del error experimental que típicamente alcanza el 20%.

En general las medidas experimentales de formación de crema concuerdan bastante bien con los resultados teóricos. En ausencia de sal, todos los sistemas son estables con respecto a gravedad durante el tiempo muy superior al tiempo necesario para estimar su constante de agregación (60 s). Destaca la estabilidad de los sistemas C y D cuya densidad se aproxima cercanamente a la densidad de la fase externa. No se determinó la velocidad de sedimentación de los sistemas en el régimen DLCA.

La Tabla 1 también muestra las constantes de Hamaker obtenidas a partir de la Ec. (18). De acuerdo con Hough y White [1980] el valor teórico de $A_H$ para dodecano es 5 x $10^{-21}$ J. Este resultado tiene el mismo orden de magnitud de los predichos por la Ec. (18). El hecho que el sistema C presente un valor sustancialmente mayor al resto, puede explicarse en base a su mayor proporción del componente más polar de las mezclas (tetracloroetileno).

En un estudio reciente [Mendoza, 2012] se evaluó la variación del radio cúbico promedio en función del tiempo para un conjunto de nanoemulsiones de composición similar a las aquí estudiadas. De acuerdo a la ecuación Ec. (7) la tasa de maduración predicha por la teoría LSW es de 1 x $10^{-26}$ m$^3$/s. De acuerdo a Mendoza y colaboradores, los sistemas, A, B, C y D muestran valores de $dR_p^3/dt$ de: 2,4 x $10^{-26}$ m$^3$/s, 1,6 x $10^{-28}$ m$^3$/s, 4,7 x $10^{-26}$ m$^3$/s, y -3,1 x $10^{-27}$ m$^3$/s. De acuerdo con estos resultados, el sistema menos estable con respecto a maduración (C) es también el sistema con mayor proporción de componente soluble (tetracloroetileno), mientras que el sistema más estable (B) es aquél que tiene la mayor cantidad de componente insoluble (escualeno). En ese mismo trabajo se evaluaron los valores de $k_{11}$ empleando la ecuación (4) obteniéndose que para salinidades mayores que 400 mM (régimen DLCA), el valor absoluto de $k_{11}$ disminuye en el orden C > D≈B ≥A, mientras que el régimen de agregación controlada por reacción (RLCA) dicha constante decrece en el orden B > A > D > C. Dado que las tasas de agregación rápida varían de un sistema a otro, el orden exhibido por el cociente de estabilidad C > D > A > B no puede predicirse fácilmente (Eq. (5)).

La Tabla 3 resume una serie de criterios comúnmente usados para establecer la estabilidad de una dispersión desde el punto de vista experimental. Como puede comprobarse, en presencia de varios fenómenos de desestabilización las tendencias son contradictorias. De los 9 criterios, el sistema C es el menos estable 6 veces, mientras que el sistema B es el más estable en 4 de ellos. Es claro sin embargo que el sistema más estable es aquél que presente la menor variación en el tiempo, algo que es difícil de deducir de la información presentada en la Tabla 3. Los distintos criterios cuantifican diversos fenómenos de desestablización ó combinaciones de los mismos. El

Tabla 2: Variación de W con la salinidad

| Sistema | $\dfrac{d \log W}{d \log[NaCl]}$ | $\dfrac{d \log W}{d [NaCl]}$ (mM)$^{-1}$ | $A_H$ (Joule) | Potencial de Stern (mV) |
|---|---|---|---|---|
| A | -5,2 ±0,3 ($r^2$ = 0,9699) | -0,0061 ±0,0005 ($r^2$ = 0,9623) | 1,6 x $10^{-21}$ | -10,9 |
| B | -3,4 ±0,3 ($r^2$ = 0,9418) | -0,0039 ±0,0005 ($r^2$ = 0,9356) | 1,1 x $10^{-21}$ | -8,9 |
| C | -19 ±2 ($r^2$ = 0,9179) | -0,021 ±0,004 ($r^2$ = 0,9204) | 5,9 x $10^{-21}$ | -21,3 |
| D | -5,2 ±0,4 ($r^2$ = 0,9461) | -0,006 ±0,001 ($r^2$ = 0,9485) | 1,7 x $10^{-21}$ | -11,3 |

Tabla 3: Criterios de Estabilidad

| Criterio | Mayor Estabilidad | Intermedia | Intermedia | Menor Estabilidad |
|---|---|---|---|---|
| Potencial Eléctrico Experimental | D | B | A | C |
| Potencial Eléctrico Ec. (19) | C | D | A | B |
| Valor Absoluto de $k_{11}$ (régimen DLCA) | A | B | D | C |
| Valor Absoluto de $k_{11}$ (régimen RLCA) | C | D | A | B |
| Valor absoluto de W a la misma salinidad | C | D | A | B |
| Valor de $A_H$ | B | A | D | C |
| CCC | 400 mM | 400 mM | 400 mM | 400 mM |
| $dR_p^3/dt$ | B | D | A | C |
| $\dfrac{d \log W}{d \log[NaCl]}$ | B | A | D | C |

Yorlis Mendoza, Kareem Rahn-Chique, German Urbina-Villalba

fenómeno dominante será aquél que genera el mayor cambio en la distribución de tamaño de gotas respecto al tiempo, algo que no puede deducirse de la información presentada. Dado que el fenómeno dominante puede cambiar como función del tamaño promedio de las gotas, los resultados evidencian la necesidad de evaluar la estabilidad de los sistemas a largo plazo ó en su defecto, emplear herramientas de predicción teórica como la estabilidad de emulsiones [Urbina-Villalba, 2009] ó los análisis numéricos del tipo Ecuación de Poblaciones a fin de predecir la estabilidad.

## 4. CONCLUSIONES

Se diseñó una estrategia experimental que permite evaluar el conciente de estabilidad de nanoemulsiones de una manera similar a como se hace con las dispersiones de partículas sólidas. Se mostraron además diversas medidas de estabilidad relacionadas con la tasa de formación de crema, la maduración de Ostwald, y la velocidad de agregación de los sistemas bajo estudio. Se concluyó que si bien el factor de estabilidad es un parámetro muy importante a la hora de definir la velocidad de agregación de las nanoemulsiones, su importancia relativa depende del fenómeno de desestabilización dominante en el sistema, el cual depende marcadamente de la composición del aceite.

Si bien es difícil establecer un orden absoluto de estabilidad para los sistemas bajo estudio, en lo que respecta a agregación y en lo que se refiere exclusivamente al efecto de la flotabilidad y la maduración de Ostwald sobre el cociente de estabilidad (Figura 6) puede concluirse que la estabilidad del dodecano puro (sistema A) se incrementa considerablemente al producir emulsiones de flotabilidad nula mediante la adición de TCE (W es mayor a la misma salinidad). Por el contrario, la estabilidad disminuye (W es menor) cuando se mezcla dodecano con escualeno (sistema B). La adición de TCE y SQ al dodecano (sistema D) genera valores de W muy cercanos a los del dodecano puro (sistema A), lo cual sugiere que el TCE lo que hace es corregir la inestabilidad producida por el SQ. Obsérvese que si mide la estabilidad del sistema a través de la variación del radio cúbico promedio de cada nanoemulsión con respecto al tiempo [Mendoza, 2012], se llega a un resultado contradictorio: la adición de SQ estabiliza sustancialmente al dodecano mientras que el TCE lo desestabiliza, lo cual se debe a la diferencia de solubilidad de éstos aditivos en agua.

Los estudios presentados fueron realizados a una concentración fija de surfactante cercana a su concentración micelar crítica (8 mM). El efecto de la concentración de surfactante sobre la importancia relativa de los fenómenos de desestabilización para cada tamaño promedio de gota merece mayor investigación.